\documentclass[twocolumn]{article}

\usepackage{amssymb}
\usepackage{adjustbox}
\usepackage{array}
\usepackage{amsmath}
\usepackage{booktabs,makecell}
\usepackage{diagbox}
\usepackage{enumitem}
\usepackage{graphicx}
\usepackage{subfig}
\usepackage{tikz}
\usepackage{verbatim}
\usepackage[round]{natbib}
\usepackage{authblk}

\DeclareGraphicsExtensions{%
	.png,.PNG,%
	.pdf,.PDF}
\usetikzlibrary{automata,positioning,arrows,trees,shapes,shapes.geometric,calc}

\newcommand*\cryssmex{\texttt{CrySSMEx}}
\newcolumntype{R}[2]{%
    >{\adjustbox{angle=#1,lap=\width-(#2)}\bgroup}%
    l%
    <{\egroup}%
}

{\def\addtocontents#1#2{}%
\def\addcontentsline#1#2#3{}%
\def\markboth#1#2{}%

\newcommand*\ttt{\texttt{t}}
\newcommand*\ttp{\texttt{p}}
\newcommand*\tts{\texttt{s}}

\oddsidemargin 6.5pc
\evensidemargin 6.5pc
\advance\oddsidemargin by -1in  
\advance\evensidemargin by -1in 
\marginparwidth 0pt             
\marginparsep 11pt              

\topmargin 5.5pc                
\advance\topmargin by -1in      
\headheight 0pt                 
\headsep 1pt                    

\advance\topmargin by -37pt     
\headheight 12pt                
\headsep 25pt                   

\textheight 630pt       
\textwidth 40pc         

\date{}
\begin{document}

\title{\textbf{SPARK: Static Program Analysis Reasoning and Retrieving Knowledge}}

\author[1]{Wasuwee~Sodsong}
\author[2]{Bernhard~Scholz}
\author[3]{Sanjay~Chawla}

\affil[1]{School of Information Technologies, University of Sydney, NSW, Australia}
\affil[2]{Department of Computer Science, Yonsei University, Seoul, Korea}

\affil[3]{Qatar Computing Research Institute, Doha, Qatar } 

\maketitle

\begin{abstract}
Program analysis is a technique to reason about programs without executing them,
and it has various applications in compilers, integrated development environments, and security. 
In this work, we present a machine learning pipeline that induces a security analyzer 
for programs by example. The security analyzer determines whether a program is either 
{\em secure} or {\em insecure} based on symbolic rules that were deduced by our machine learning pipeline.  The machine pipeline is two-staged consisting of
a Recurrent Neural Networks (RNN) and an Extractor that converts an RNN to symbolic rules. 

To evaluate the quality of the learned symbolic rules, we propose a sampling-based similarity measurement between two infinite regular languages.  We conduct a case study using real-world data. In this work, we discuss the limitations of existing techniques and possible improvements in the future.  The results show that with sufficient training data and a fair distribution of program paths it is feasible to deducing symbolic security rules for the OpenJDK library with millions lines of code.

\end{abstract}

\section{Introduction}
The aim of static program analysis (SPA) is to reason about the behavior of computer programs without actually
running them. One motivation for SPA is encapsulated in the famous saying of Djikstra, ``Program testing
can be used to show the presence of bugs but never their absence.'' In this paper we show that we can use
machine learning (in particular RNNs) to deduce symbolic rules for program analyzers. We would like to learn symbolic rules that decide whether a program is either {\em secure} or {\em insecure} according to some security coding guidelines. The symbolic rules are an approximation for the concrete program semantics of a given program. The generated program analyzer will have a high but not perfect
accuracy since sound and complete program analysis is an undecidable problem. 

Our machine learning pipeline for inducing a program analyzer consists of the following components: (i)~a
database of programs  (ii)~representation of a program as an abstract state machine (ASM) that condenses a program into a state machine reflecting the semantic properties that are relevant for the security analysis. Reading along the edges of the ASM each program path is represented as a sequence of symbols 
and  labeled as {\em secure} or {\em insecure}; (iii)~the use of recurrent neural networks (RNNs) to build a 
binary classifier using the labeled paths; (iv)~the extraction of a probabilistic finite state machine from the trained RNN for generating interpretable rules for the determining the security status of a program.

\begin{figure}
	\includegraphics[width=\linewidth]{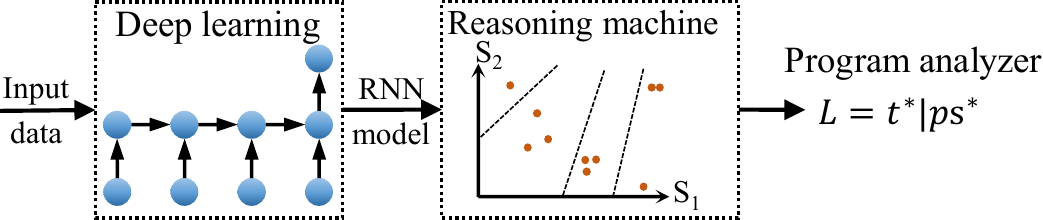}
	\caption{Program analyzer induction pipeline.\label{fig:overview}}
\end{figure}

Knowledge extractions from RNNs were heavily studied in the 1990s~\citep{cleeremans89,giles92,manolios94,omlin96}. Due to computational limitation at the time, the problems were applied to  simple recurrent networks on small 
languages with few neurons.  The latest rule extraction framework was proposed by Jacobsson in 2006~\citep{jacobsson06}.  In contrast, the study of neural networks (under the deep learning rubric)  has recently received 
renewed interest. 
In particular, long short-term memory (LSTM) networks are known to resolve the vanishing gradient program in 
previous RNN models enhancing ability of learning long-term dependencies. Nevertheless, no previous study has attempted extract rules from the network.  Our framework couples a long short term memory (LSTM) network with 
Jacobsson's \cryssmex{}~\citep{jacobsson06}. 

To show the effectiveness of the produced symbolic rules, we compute the similarity
between two symbolic rule systems with the notion of edit distance. This similarity
metric is necessary for measuring the quality of the learned rules. Assume we have a set of sentences $S$ from a language $L$. Let $\hat{L}$
be the language induced (using an RNN) from $S$. Then how do we compare $L$ and $\hat{L}$? 
Similarity metric comparing two infinite regular languages was proposed by~\citet{benedikt11}.
However, the computation of the metric requires double exponential time.  We propose a sampling-based 
similarity comparison that computes the lower bound in polynomial time complexity.

Our contributions are as follows:
\begin{itemize}[noitemsep,,topsep=0pt]
	\item[$\bullet$] We propose the first formulation of a static-analysis problem in a machine learning framework.
	\item[$\bullet$] We use modern deep learning methods for extracting finite state automata.
	\item[$\bullet$] We propose a heuristic polynomial time similarity metric of two infinite regular languages.
	\item[$\bullet$] We have carried out a detailed experimental study on openJDK, consisting of millions of lines of code.
\end{itemize}

The remaining of this paper is structured as follows. We first demonstrate a motivating example in Section~\ref{sec:motivation}. Then, we explain an overview and detailed implementation in Section~\ref{sec:overview} and Section~\ref{sec:method} respectively. 
The details of our lower bound similarity metric are given in Section~\ref{sec:similarity}. In experimental results and evaluations are presented in Section~\ref{sec:experiment}.
Related work is discussed in Section~\ref{sec:relwork}; we conclude in Section~\ref{sec:conclusion}.

\section{Motivating Example}
\label{sec:motivation}

\begin{figure}[tb]
	\begin{center}
		\begin{tikzpicture}[ state/.style={draw, circle, fill=white, 
			minimum size=1.5em}, 
		scale=0.8, every node/.style={scale=0.8}]
		\matrix[row sep = 0.5em,column sep=0.5em] {
			& \node[state] (a1) {0}; & & \node {\hspace{0.5em} }; & \node[state] (b1) {0}; & & \node {\hspace{2em} }; & &\node[state] (c1) {0};& \\
			
			\node[state] (a2) {1}; & &\node[state] (a4) {3};& &\node[state] (b2) {1};& & &\node[state] (c2) {1};& &\node[state] (c4) {3};\\ 
			
			\node[state] (a3) {2}; & & & &\node[state] (b3) {2};& & &\node[state] (c3) {2};& &\node[state] (c5) {4};\\ 
			
			& \node[state,accepting, double distance=1pt] (a5) {4}; & & & \node[state,accepting, double distance=1pt] (b4) {3}; & & & & \node[state,accepting, double distance=1pt] (c6) {5};& \\
		};
		\node[above left=0.3 cm of a1](i1){};
		\node[above left=0.3 cm of b1](i2){};
		\node[above left=0.3 cm of c1](i3){};    
		\path [-latex]
		(i1) edge (a1)
		(i2) edge (b1)
		(i3) edge (c1)      
		(a1) edge node [above, pos=0.7] {\ttp} (a2)
		(a2) edge node [left] {\ttt} (a3)
		(a3) edge node [below, pos=0.3] {\tts} (a5)
		(a1) edge node [above, pos=0.7] {\ttt} (a4)
		(a4) edge [loop right]  node [above] {\ttt} (a4)
		(a4) edge node [right] {\ttt} (a5)
		(b1) edge node [left]  {\ttt} (b2)
		(b2) edge node [left]  {\ttp} (b3)
		(b3) edge node [left]  {\tts} (b4)
		(c1) edge node [above, pos=0.7] {\ttp} (c2)
		(c2) edge node [left]  {\tts} (c3)
		(c3) edge node [below, pos=0.2]  {\ttt} (c6)
		(c1) edge node [above, pos=0.7] {\color{red}\tts} (c4)
		(c4) edge node [right]  {\ttt} (c5)
		(c5) edge node [below, pos=0.2]  {\ttt} (c6)   
		;
		\node [below= 1.5em of $(a5)!0.4!(b4)$] (l) {\large{(a) secure programs}};
		\node at (c6 |- l) {\large{(b) insecure program}};
		\end{tikzpicture}
		\vspace{-3mm}
	\end{center}
	\caption{Finite state machine examples of two secure programs and one insecure program. \label{fig:securityfsa}}
\end{figure}
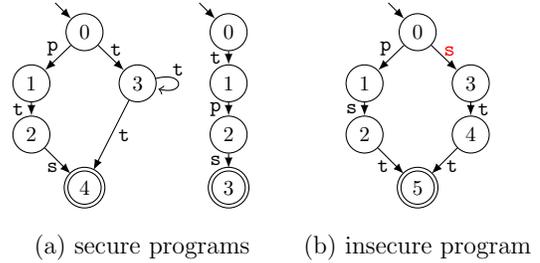

This section provides an intuitive example of learning security analyzer by example.
We demonstrate an induction of a security rules which can be used to classify secure and insecure programs. 
A program is \textit{secure} if there exists at least one permission check before a security call; otherwise it is \textit{insecure}. The question is whether such a symbolic rule can be learned passively via examples. 
An example of two secure programs and one insecure program in form of abstract state machines are illustrated in Figure~\ref{fig:securityfsa}.
The state machines have three types of labels: permission check \ttp,  security-relevant call \tts, and other (\textit{transparent}) statement \ttt. 
A program is secure if on all paths from the start state, $0$, to all accepting states there exists at least one permission check, \ttp, before every security call, \tts.
Hence, the program in Figure~\ref{fig:securityfsa}(b) is insecure because there exists a security-relevant call without a permission check on the LHS path.

\begin{figure}[tb]
	\begin{center}
		\begin{tikzpicture} [
		nnode/.style={draw, circle, fill=white, minimum size=1em}, 
		mat/.style={row sep = 1em,column sep=1em}, scale=0.8, every node/.style={scale=0.8}
		]
		\matrix[mat] (A){
			& \node [nnode,double, double distance=1pt, align=left] (a0) {0}; &\\
			\node [nnode,double, double distance=1pt, align=left] (a1) {1}; & & \node [nnode, align=left] (a2) {2}; \\
		};
		\node [above left = 0.3 cm of a0] (initA){};
		\path[-latex]
		(initA) edge (a0)
		(a0) edge node [left,pos=0.4]{\ttp} (a1)
		(a0) edge node [right,pos=0.4]{\tts} (a2)
		(a0) edge[loop right] node {\ttt} (a0)
		(a1) edge[loop left] node {\ttt,\tts,\ttp} (a1)
		(a2) edge[loop right] node {\ttt,\tts,\ttp} (a2);  
		\end{tikzpicture}
		\caption{The target finite state machines of our security analysis.
			\label{fig:targetfsa}}
	\end{center}
\end{figure}
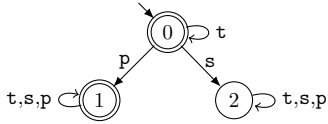

Labels along paths in the state machine form strings that describe behaviors of program paths.
In this example, the secure paths are \texttt{pts, tps} and \texttt{tt}$^*$\texttt{t}.
The only insecure path among the three programs is `\texttt{stt}'. 
Combining knowledge from these limited examples, the secure behavior maybe induced as a program that does not begin with the security call, \tts.
This rule only holds true for the particular examples but may not hold true for all programs.
With more examples, the more generalized rules can be derived.
The regular expression of this security analysis is $\mathcal{L}~=~\ttt^*|\ttp(\ttt|\ttp|\tts)^*$. The target FSA of our security analysis is shown in Figure~\ref{fig:targetfsa}.

The security analysis can be reshaped as a regular language reachability problem~\citep{reps98}. 
Given a regular language $\mathcal{L}$, a program path is secure if a string equivalent is in $\mathcal{L}$. However, $\mathcal{L}$ is not known ahead of time.
In this paper, we propose a machine learning pipeline that automatically induces the symbolic rules, representing as the language $\mathcal{L}$, from user examples.

By tracing along the edges of the state machine, sets of strings representing secure and insecure behavior can be generated.
We train an RNN model as a binary classifier to detect secure and insecure program paths. 
However, the black-box learning model is not understandable to users 
and we wish to extract symbolic rules by a reasoning machine 
(or a rule extractor). 
A rule extraction framework takes an RNN model as its input and performs state space quantization generating an FSA counterpart of the RNN model. 
This FSA is the security analyzer, a meta-program that can determine whether a programs is secure, thus, making the property of secure program understandable to users.

\section{System Overview}
\label{sec:overview}

\begin{figure}[tb]
	\begin{center}
		\subfloat[Practical setup]{\includegraphics[width=0.8\linewidth]{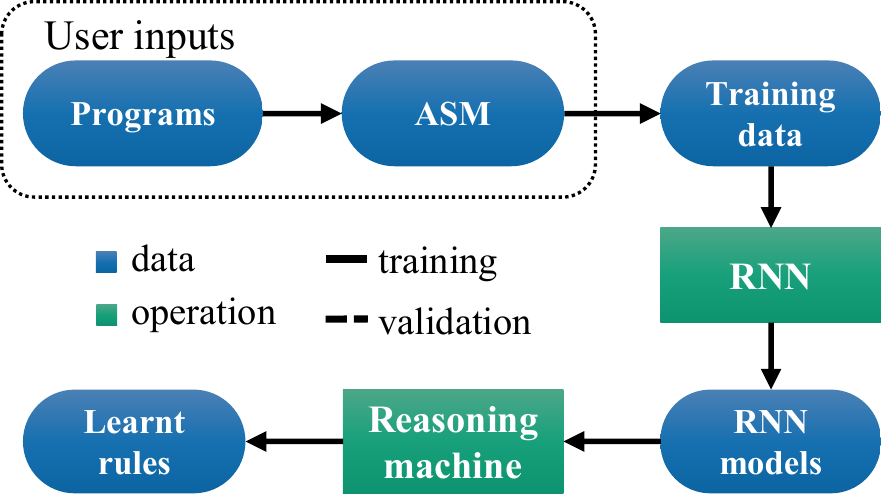}}\\
		\subfloat[Experimental setup]{\includegraphics[width=0.8\linewidth]{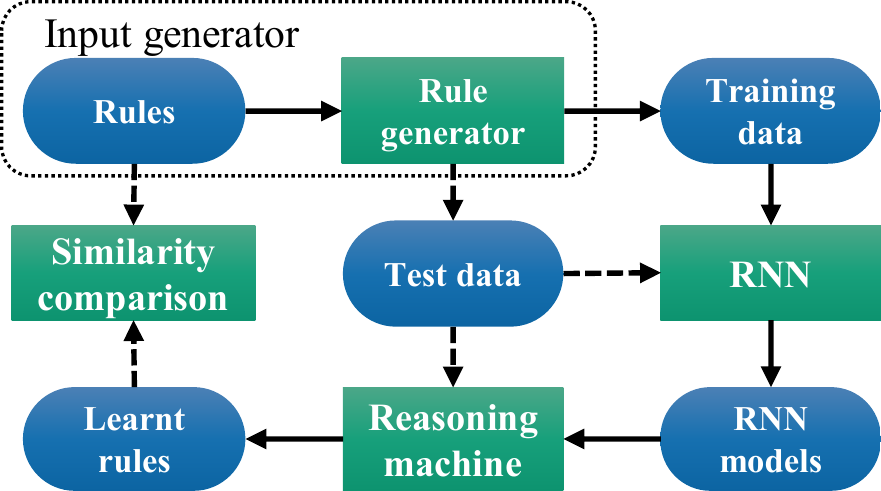}}		
	\end{center}
	\vspace{-3mm}
	\caption{The (a) practical setup and (b) experimental setup of the program analyzer induction pipeline.\label{fig:pipeline}}
\end{figure}

The overview of our inductive reasoning pipeline using recurrent neural networks is depicted in Figure~\ref{fig:pipeline}~(a). 
Neural network is unable to take user programs as input directly.
Programs often contain excessive information to the target analysis which may dominate the important features during training. Hence, they must be converted to abstract state machines (ASMs) removing inessential information to the analysis. The remaining relevant information is then encoded as transition labels on the ASM.
Hence, tracing ASM forms a set of strings describing behavior of the program with respect to the analysis.
The program paths in strings and their classifications as positive or negative examples are used as an input dataset to RNN. 
We model RNN as a binary classifier that predicts whether execution paths are accepted or rejected by the analysis.
We consider a binary classifier because our symbolic rules, represented as a regular language, have two output states: accepting and rejecting.
Through the training, we expect the rules to be captured within the learned RNN model. 
However, the model may not be understandable to users. 
We apply a rule extractor as our reasoning machine, which quantize the continuous state space of RNN to a discrete state machine.
The output rules, in a regular expression, derives a new and reusable program analyzer.

In practice, customizing ASM labels and classifying validity of program paths are subjective to the analysis aspects. 
Automated classifications of program paths are a challenge problem by itself and is not covered in this paper.
Hence, the decisions should be made by users. 
To evaluate, we consider a closed experimental setup where the program analyzer is known ahead-of-time. 
Consequently, we replace user inputs with our input generator as depicted in Figure~\ref{fig:pipeline} (b). The input generator randomly generates dataset according to the pre-defined rules.
As the result is known beforehand, the soundness of learning can be evaluated using similarity measurement.
Because there exists no solutions that compare two different grammars in reasonable computable time~\citep{higuera10}, we propose a similarity metric that calculate a lower bound editing ratio between two regular languages in Section~\ref{sec:similarity}.

\section{Inductive Reasoning}
\label{sec:method}

In this section, we explain each process our static analyzer induction pipeline in greater details. 

\textbf{Abstract State Machines: }
The source programs represented as control flow graphs may capture irrelevant information to the analysis. For instance, only permission checks and security calls are applicable to the security analysis.
We simplify information by converting to a more concise representation, namely an abstract state machine (ASM).
Unlike the conventional CFG, an ASM encodes basic blocks as labels on the edge. 
Only basic blocks of interests are labeled. The information of other basic blocks are abstracted out and considered as transparent edges,~\ttt.
The labels are problem specific, and thus, are specified by users.
In the security analysis, we are only interested in permission check and security call statement; hence, the labels are \ttp~and \tts, \ttt. 
Traversing along an ASM generates strings representing program paths. 
It should be noted that there may be an infinite set of strings due to loops. 
We sample strings from all ASMs and used as an input to our RNNs. 

\textbf{Recurrent Neural Networks:}
Rule extraction from RNNs (refer as RNN-RE) has been actively studied in the 1990s~\citep{giles92,cleeremans89,watrous91,omlin96,giles89,manolios94}.
With the computational limitation at the time, the rules were extracted from simple recurrent networks with few hidden states. 
The networks are trained with a handful of training data, and repeatedly trained in a high number of iterations. 
Moreover, early RNN models have a problem of vanishing gradients; as a result, the influence of early terms are overshadowed by the later terms in a long sequence. 
With recent advent of neural networks, we choose to use a long short-term memory (LSTM) and gated recurrent unit (GRU), a variation of LSTM, as our hidden layer than they are able to capture long-term dependency. 
This is the first work that applies rule extraction on LSTM/GRU models. 
We implement our neural network with three layers: input, LSTM/GRU and output layer. 
One character is fed to the network at a single time sequence. 
The binary classifier predicts whether a string ends in an accepting state; thus, the output layer has a dimensionality of one. 
The input layer, $I$, is a binary vector of $L$ symbols, where $|\Sigma| = L$. At most one index of the input vector can have the value of one.
The input layer is connected to an LSTM layer, $S$ of $N$ neurons.
Only the hidden layer of the last time step, $S_t$, is used to predict the outcome. We discuss the configuration of the RNN and its trade-off with respect to rule extractions in more details in Section~\ref{sec:experiment}.

\textbf{Reasoning Machine: }
RNN can be described similar to a Mealy machine~\citep{jacobsson05} with a hidden state transition and output transition from a given input. 
In contrast, RNN states are in continuous space and has computational power of Turing equivalent~\citep{siegelmann92}. 
The hypothesis behind FSA extraction from RNN is that after sufficient network training, the network's state space clearly diverges, and clusters of state vectors correspond to a state of an FSA. 
The rule extraction commonly share the following steps: generating hidden states and outputs by feeding input patterns, clustering and discretize the continuous state space of RNN to FSA, establish state transitions and minimize FSA states. 
A comprehensive survey and compilation of open issues of the early works are summarized in~\citep{jacobsson05}. 

The early rule extractions give equal weight to geometric properties of the state space. However, the geometric distance does not directly reflect the impact to the network as a result of feedback loops within the network.
Jacobsson~\citep{jacobsson06} proposed Crystallising Substochastic Sequential Machine Extractor (\cryssmex{}) which takes into account this property. 
Given RNN's inputs, internal states and outputs, \cryssmex{} extracts a probabilistic finite state machine.
Initially, \cryssmex{} constructs a machine, $\mathcal{M}$, with the entire state space as a single non-deterministic FSM state. At each iteration, it first selects the most non-deterministic states based on the probability. Then state quantizer is updated by splitting the RNN state space of the corresponding states. The machine $\mathcal{M}$ is updated with the new state quantizer, before merging equivalent states.

The state quantizer arranged in tree structure where each node and its level describes split region in the state space.
The extraction process terminates when all states in $\mathcal{M}$, are deterministic, or  an iteration limitation is reached. The state machine $\mathcal{M}$ represents the learned rules extracted from RNN.

\section{Similarity Metric}
\label{sec:similarity}
The quality of learned knowledge is imprinted with propagated errors from approximations in both RNN modeling and discretization in the rule extraction. 
The imprecisions in RNN are largely caused by three factors: a size of training examples,
a size of hidden layers and a duration of training.
Rule extractions cluster the continuous state space of RNN models. The computational power of the network is reduced from Turing-equivalent to a finite state machine and thus, it propagates further errors to the output FSA. 
The error introduced throughout the program analyzer induction pipeline may constitute in the learned FSA. None of the previous studies in rule extractions have comparison metrics to evaluate the learned rules.

In order to evaluate the quality of learning, we construct a control setup (Figure~\ref{fig:pipeline}(b)) where inputs are generated from known rules. In this control setup, the quality of learning can be evaluated by comparing the similarity between the original rules and the learned rules expressed as infinite regular languages.

\citet{benedikt11} proposed an algorithm calculates the normalized traveling cost between languages with infinite strings. However, the algorithm is computable in double exponential time complexity. 
We propose a similarity metric that calculate a lower-bound cost of traveling between the learned FSA and the target FSA.  
Given two regular languages $L_A$ and $L_B$, the traveling cost is defined by
\begin{equation}
\delta(L_A, L_B) = max \left\{\frac{\texttt{edit}(w,L_B)}{|w|} : \forall w \in L_A\right\}.
\end{equation}

We extend the edit-distance calculation between a word and a regular language~\citep{wagner74} to the traveling cost between two regular languages. The traveling cost is the maximum edit ratio from words in $L_A$ to $L_B$.
The lower editing cost reflects higher similarity, and the edit cost of zero indicates that $L_B$ contains all strings of $L_A$ (i.e.,  $L_A \subseteq L_B$). Hence, we refer to this ratio as dissimilarity ratio.
It is importance to note that the edit-distance between two languages is not symmetric. 
Considering $L_A \subset L_B$, there exist words that accepted by language $L_B$ but not $L_A$. Thus, $\delta(L_A, L_B) < \delta(L_B, L_A)$ 
Hence, our dissimilarity ratio is formulated as follows:
\begin{equation}
\Delta(L_A,L_B) = max ( \delta(L_A, L_B),
\delta(L_B,L_A) ).
\end{equation}
The range of dissimilarity ratio is $[0,\infty]$. When $\Delta(L_A,L_B)$ is zero, the two languages are equivalent. The dissimilarity ratio can be above one when the number of edits is larger than the length of the given word. It should be noted that the ratio of one does not necessary mean that the entire word must be edited. Considering $L_B = (abc)^{*}$, transforming $aa \in L_{A}$ to $L_{B}$ requires two edits for replacing the second $a$ with $b$ and inserting $c$. Thus, it has dissimilarity ratio of 1. 

It is impossible in practice to compare two infinite languages. 
We approximate the lower-bound by taking word samplings from language $L_A$. 
\begin{equation}
\widehat{\delta}(L_A, L_B) = max \left\{\frac{\texttt{edit}(\widehat{w},L_B)}{|\widehat{w}|} : \forall \widehat{w} \in \widehat{L_A} \subseteq L_A\right\},
\end{equation}
Since $\widehat{\delta}(L_A,L_B) \le \delta(L_A,L_B)$, $\widehat{\delta}(L_A,L_B)$ is the upper lower bound of the travelling cost.

Editing one word, $w$, to a language with $Q$ states has a time complexity of $O(|w| \cdot |Q|^3)$. Our lower-bound cost computation operates on sample strings, and thus, has the complexity of $O(|\widehat{w}| \cdot |Q|^3 \cdot |\widehat{L_A}|)$.
To avoid misinterpretation in  lower-bound comparison, we generate a large set of strings to examine the similarities.  For each language, we generate 10,000 unique strings.

This metric is only suitable to evaluate learned FSA to one target regular language. The edit ratio should not be used to compare language similarity of two different problems.
A language with a longer shortest non-accepting string is likely to have lower ratio. For example, a regular grammar accepts strings that do not contain three consecutive zeros. The shortest non-recognisable string is \texttt{000} of length three. The maximum edit ratio of this language is $\frac{1}{3}$ by editing one of \texttt{0}'s to \texttt{1}. On the other hand, in the security analysis, \tts~is the shortest incorrect path. Hence, the maximum ratio is 1. 
Therefore, the edit-ratio cannot be used to cross-evaluate the similarities of two different problems.

The necessity of similarity metric in a control setup is self-evident. 
In the practical setup (Figure~\ref{fig:pipeline}(a)), where the analysis rules are not known, the similarity metric can be employed to select the best representative rules. 
Training RNN with various configurations produces different models as inputs to the rule extractor and may eventually induce different symbolic rules.
Regardless of differences in learning, all results are induced from the same set of execution traces, and they attempt to express the same knowledge captured in RNN models.
Hence, we consider the best representative rule as a language that has the highest similarity to all other learned rules. 

We conduce a pairwise comparison matrix among all learned rules where each row represents dissimilarity ratio of a rule to all others. The mean value of each matrix row represents an average similarity of a solution.
The best representative rules is the rule with the lowest average dissimilarity ratio (i.e., the highest similarity).

\begin{figure}[tb]
	\captionsetup[subfloat]{labelformat=empty}
	\begin{center}
		\bgroup
		\setlength\tabcolsep{2pt}
		\renewcommand{\arraystretch}{0.9}
		\begin{tabular}{c| c c c c}
			\hline 
			\diagbox{N}{Ep}
			& 500 & 1000 & 3000 & 5000 \\\hline  \hline
			4 &
			\subfloat[\scriptsize{$\Delta = 0.25$}]{\includegraphics[width = 0.44 in, height=0.85 in,clip]
				{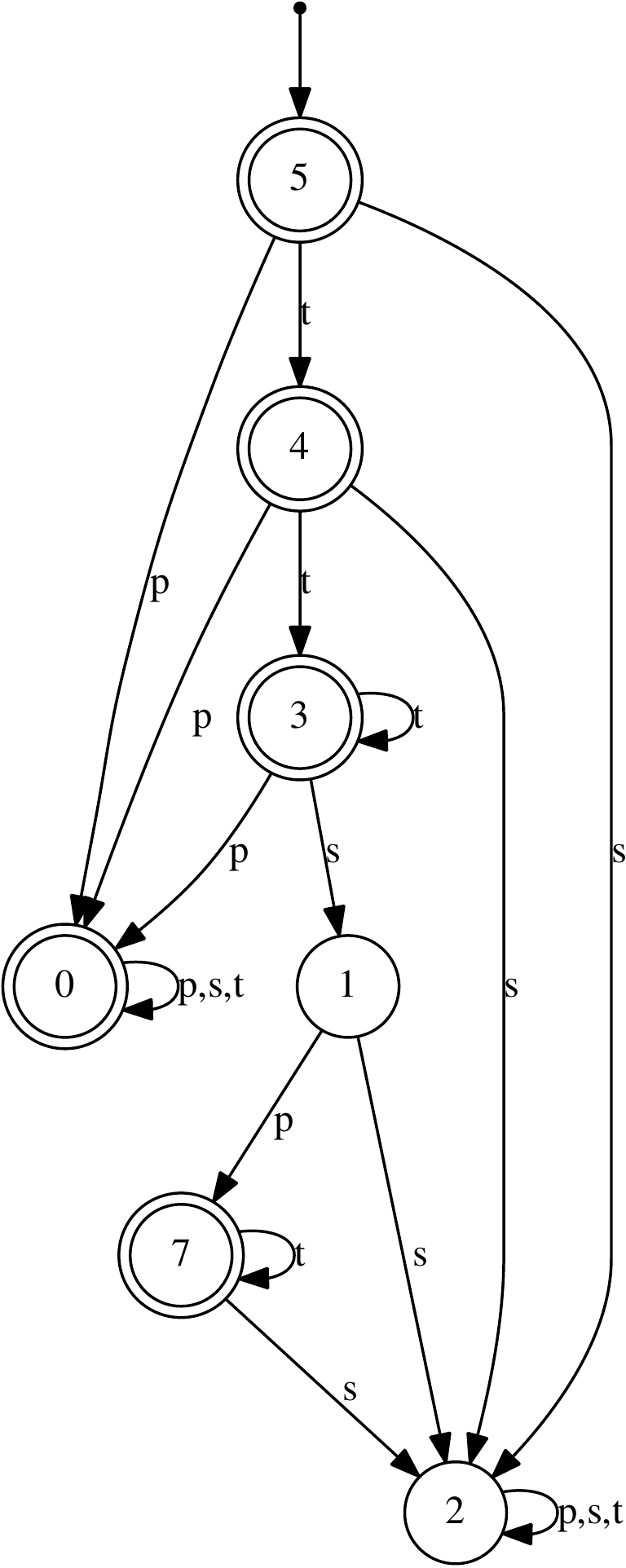}} &
			\subfloat[\scriptsize{$\Delta = 0.20$}]{\includegraphics[width = 0.45in, height=0.8 in]
				{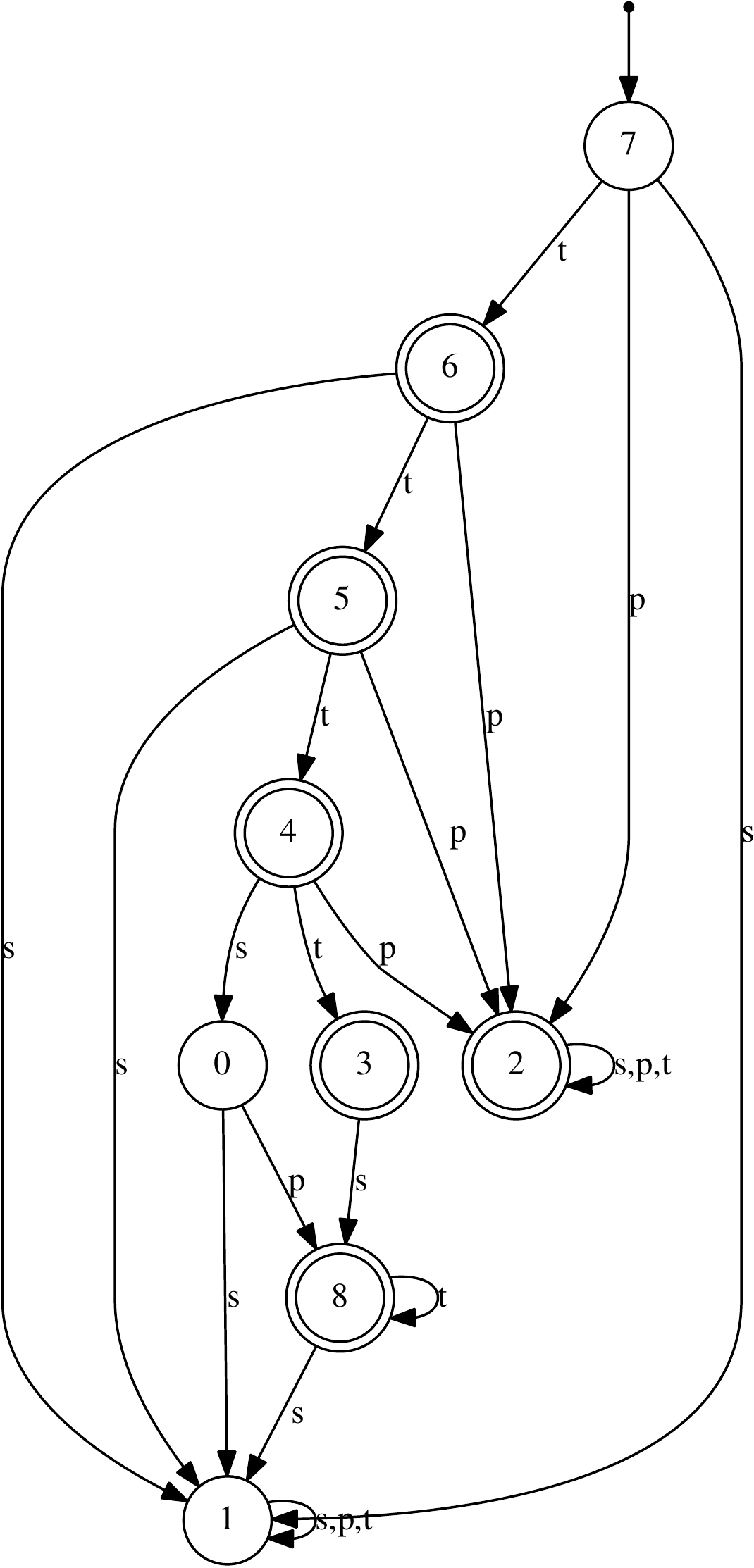}} &
			\subfloat[\scriptsize{$\Delta = 0.00$}]{\includegraphics[width = 0.7 in]
				{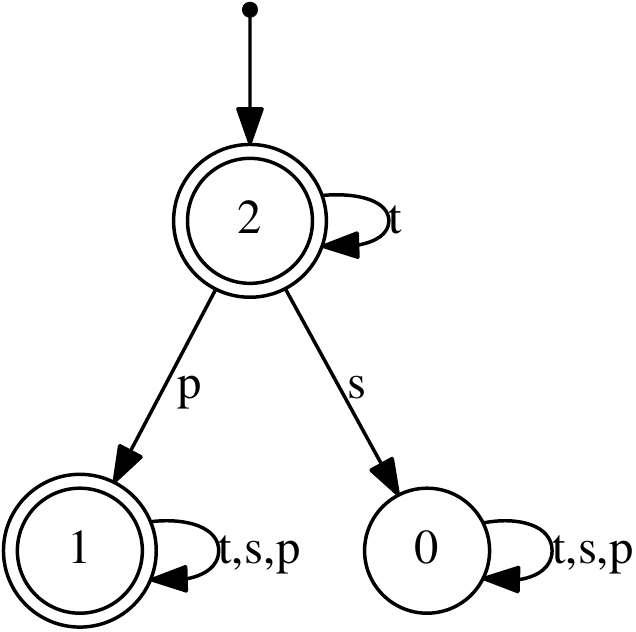}} &
			\subfloat[\scriptsize{$\Delta = 0.00$}]{\includegraphics[width = 0.7 in]
				{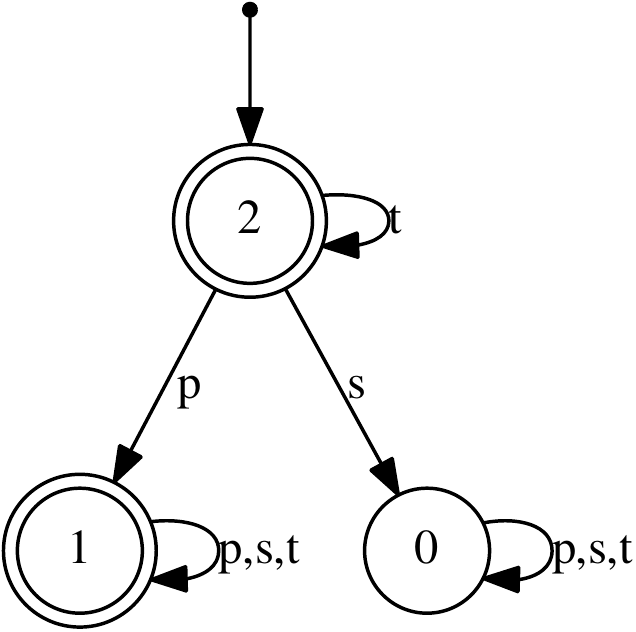}}\\ \hline
			
			8 &
			\subfloat[\scriptsize{$\Delta = 0.25$}]{\includegraphics[width = 0.45in, height=0.8 in] 
				{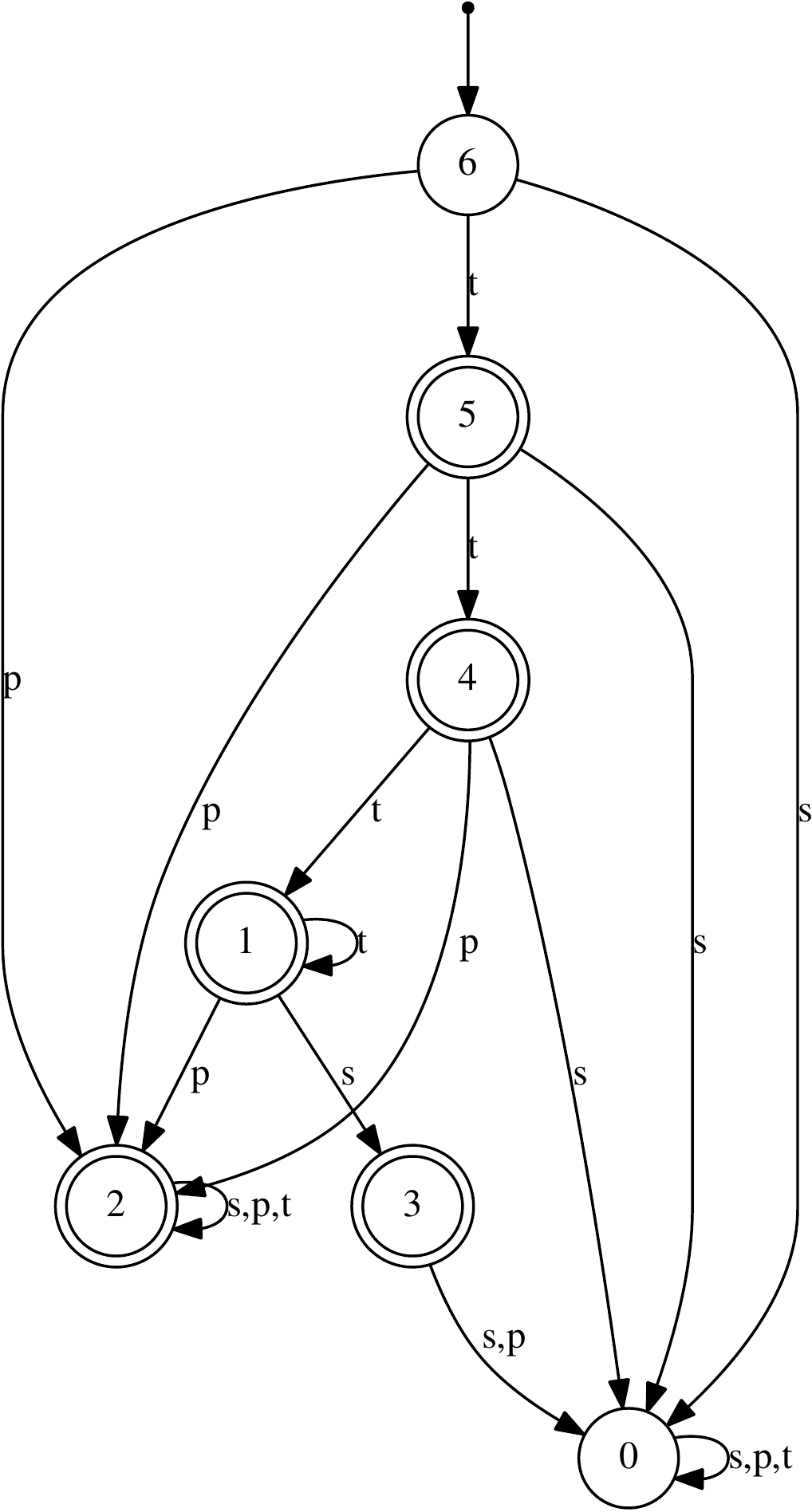}} &
			\subfloat[\scriptsize{$\Delta = 0.25$}]{\includegraphics[width = 0.45in, height=0.8 in]
				{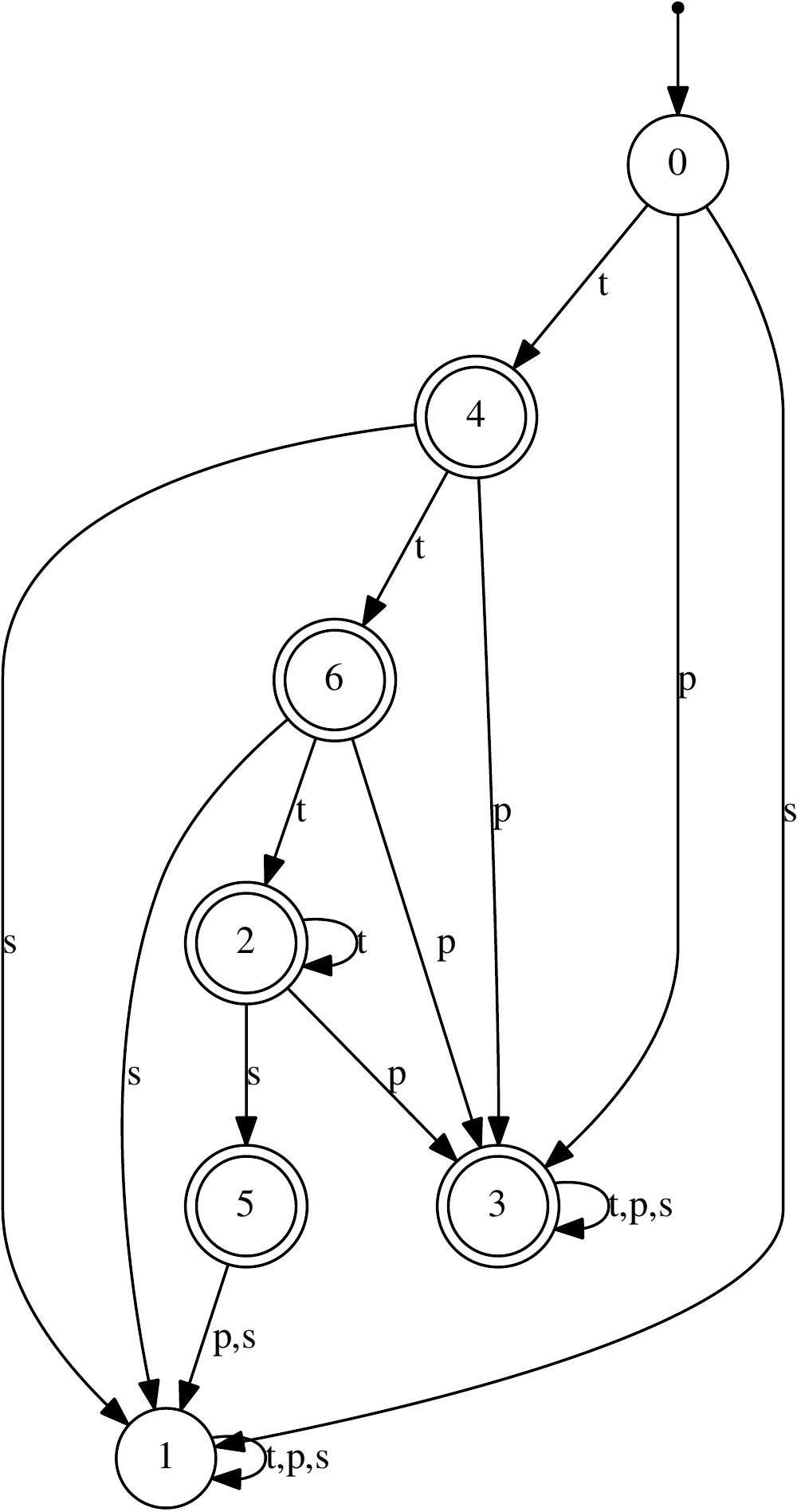}} &
			\subfloat[\scriptsize{$\Delta = 0.00$}]{\includegraphics[width = 0.7 in] {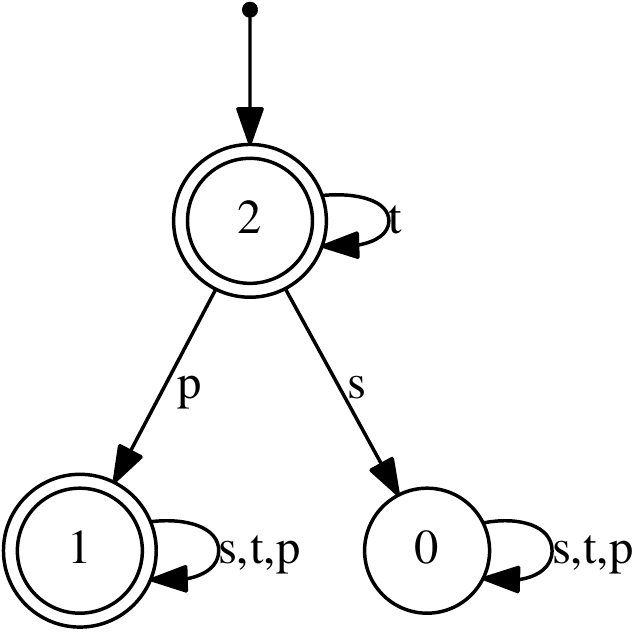}} &
			\subfloat[\scriptsize{$\Delta = 0.00$}]{\includegraphics[width = 0.7 in] {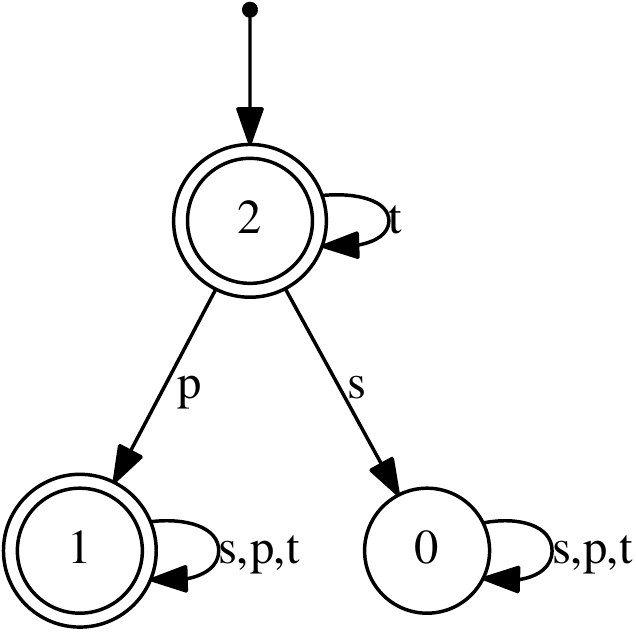}}\\  \hline
			
			16 &
			\subfloat[\scriptsize{$\Delta = 0.00$}]{\includegraphics[width = 0.7 in] {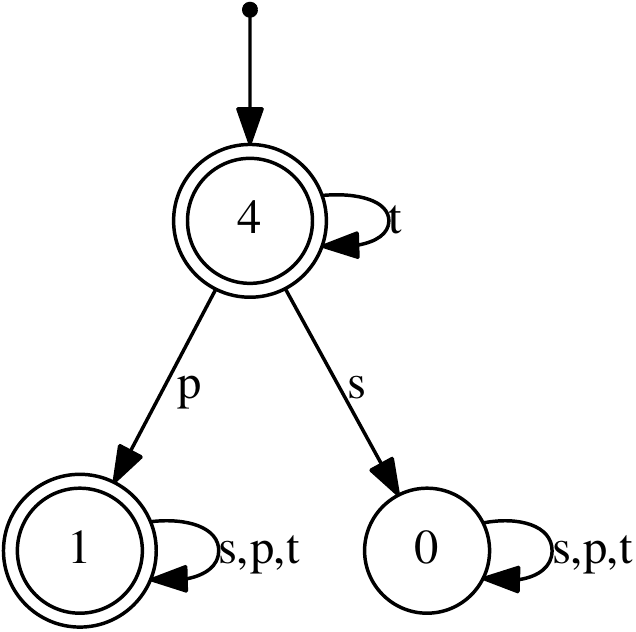}} &
			\subfloat[\scriptsize{$\Delta = 0.50$}]{\includegraphics[width = 0.45in, height=0.8 in] {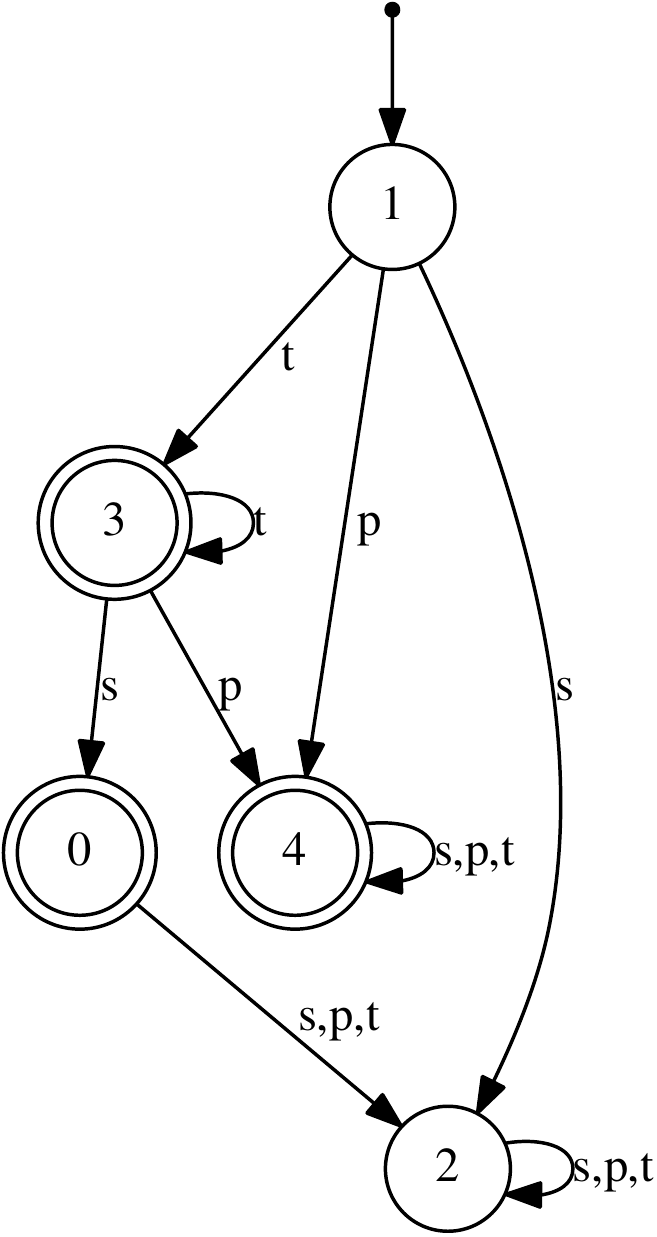}} &
			\subfloat[\scriptsize{$\Delta = 0.50$}]{\includegraphics[width = 0.45in, height=0.75 in] {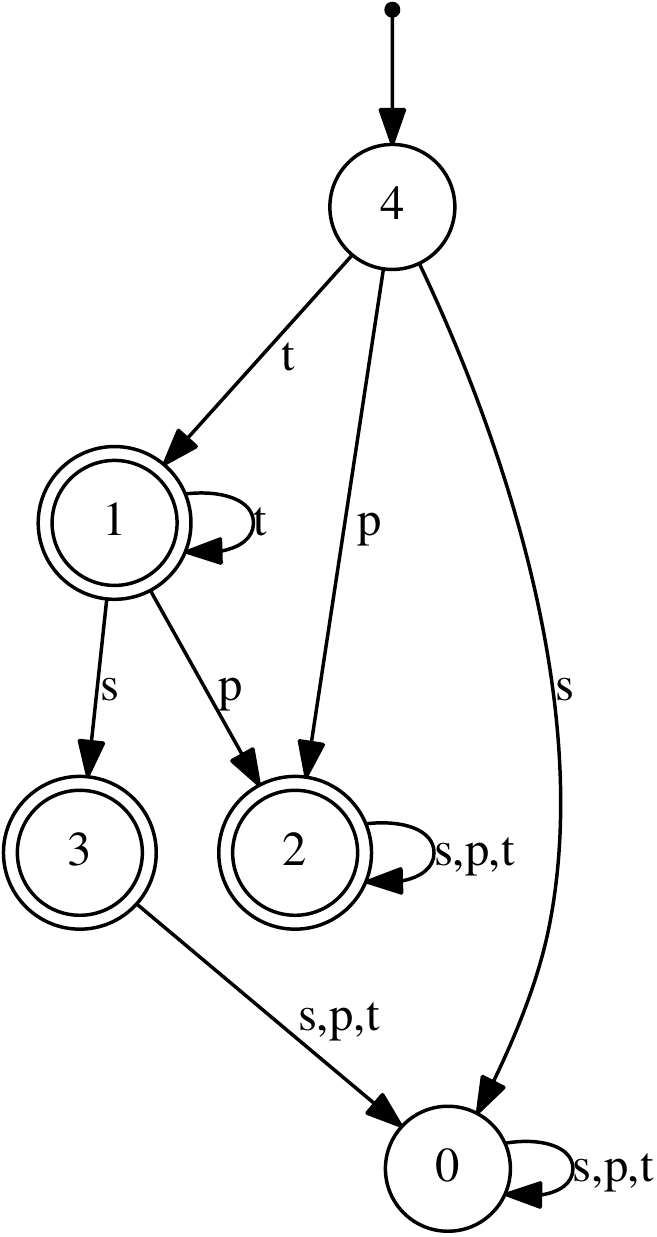}} &
			\subfloat[\scriptsize{$\Delta = 0.50$}]{\includegraphics[width = 0.45in, height=0.75 in] {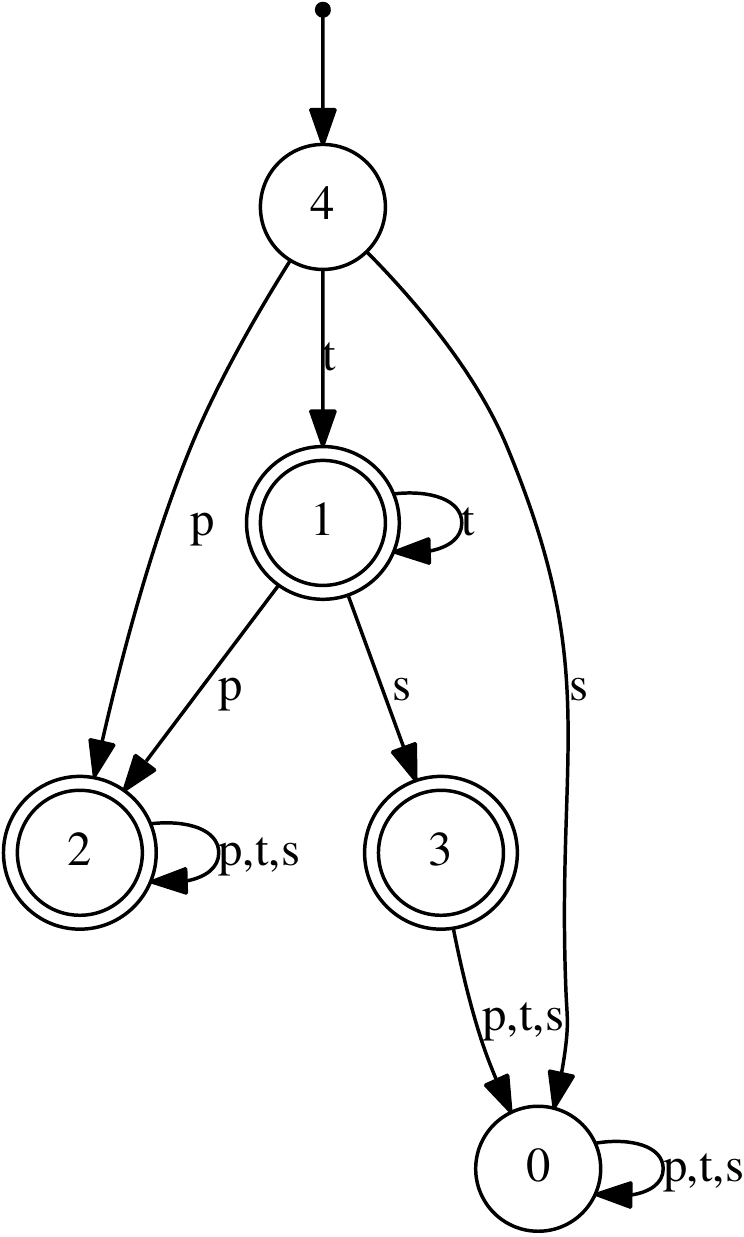}}\\
			\hline
		\end{tabular}
		\caption{Rule extraction of the security analysis example as variations of neural network sizes, N, and a number of epochs, Ep. \label{fig:learnedfsa}\vspace{-3mm}}
		\egroup
	\end{center}
\end{figure}

\section{Case Study}
\label{sec:casestudy}
\label{sec:experiment}
We are interested in the quality of induction with respect to neural network configurations, especially, the minimal learning requirements to retrieve knowledge.
For evaluation purpose, we conduct closed environment experiments where the symbolic rules are known ahead of time.
Our symbolic rules, in form of FSA over $\Sigma = \{\ttt, \ttp, \tts\}$, is depicted in Figure~\ref{fig:targetfsa}. 
In the experiments, we induce the security analyzer from synthetic inputs and then extend to real-world datasets.

In RNN training, we configure the neural network as follows.
The hidden layer contains 4 to 32 neurons. 
The output layer of our neural network is a binary value representing accepting and rejecting state; hence, we use a binary cross-entropy (BCE) as our error criterion. Each training runs up to 5000 epochs. 
For verification, the learned models are evaluated with 10,000 random strings.

We use \cryssmex{} framework as our reasoning machine to extract FSA from an RNN. 
This rule extraction framework recognizes an initial state as function to view the domain, and thus does not consider the initial state in the computation.
We assign the initial state as a state such that achieves the highest accuracy on the training data.
\cryssmex{} terminates when a deterministic solution is found. In practice, the deterministic solution may not be found in reasonable time.
Hence, we retrieve a non-deterministic solution by restricting the maximum execution iterations.

\subsection{Synthetic Experiment}
We randomly generate secure and insecure input strings from the target FSA following Figure~\ref{fig:targetfsa}. The generated strings are unique and divided into a training set and a test set.
As RNN has proven much success in natural language processing, it is not surprising that RNN models achieve very high accuracy with few neurons.
However, the rule extraction is very sensitive to correctness in RNN models. Overfitting and underfiting may lead to unsuccessful induction.
We demonstrate the fact by training RNN with 800 positive and negative examples.
Figure~\ref{fig:learnedfsa} shows learned FSAs with respect to a number of epochs and sizes of the hidden layer.
The extractions at 500 and 1,000 epochs with 4 and 8 neurons are  incorrect due to underfitting. 
On the other hand, the rule extractor failed to induce the rules correctly at 16~neurons and over 500~epochs as a result of overfitting. RNN models memorize  patterns of training examples instead of induce the solution.
The results show that the correct solutions can be extracted from particular network setups. 
Reduction of overfitting effects potentially improves the language extraction.

\begin{figure}[tb]
	\begin{center}
		{\includegraphics[clip]{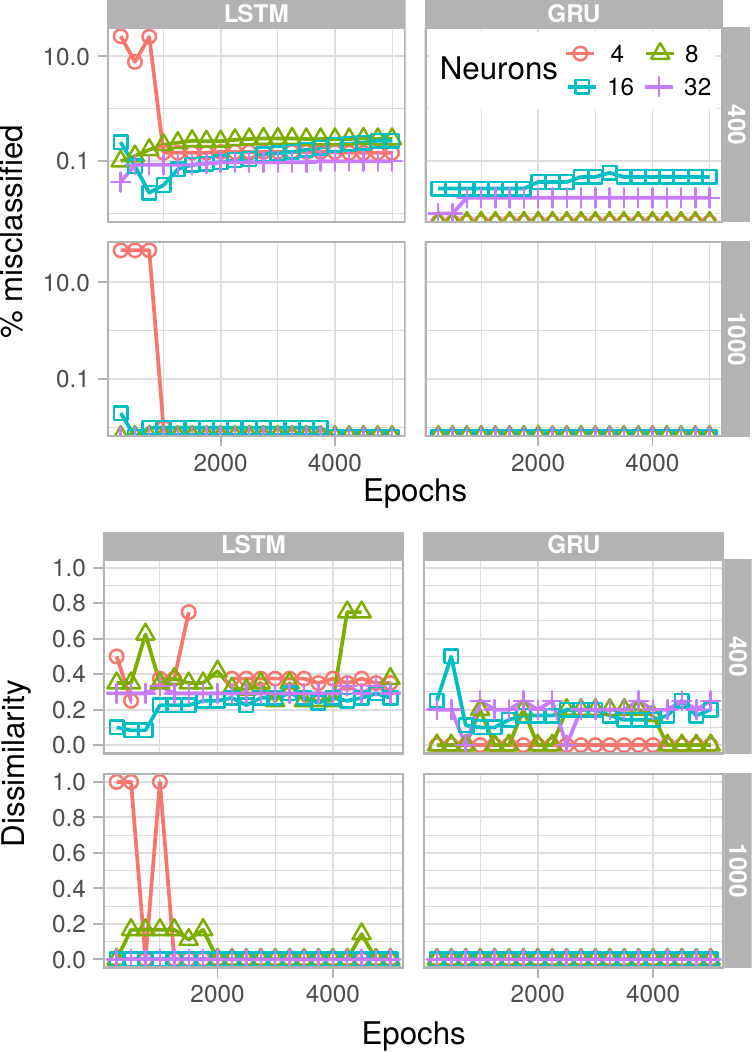}} \\
		\caption{LSTM and GRU network comparisons with 400 and 1000 training data set. GRU models show lower misclassifications and dissimilarity ratios.\label{fig:lstmgru}\vspace{-3mm}}
	\end{center}
\end{figure}

\begin{figure}[tb]
	\begin{center}
		{\includegraphics{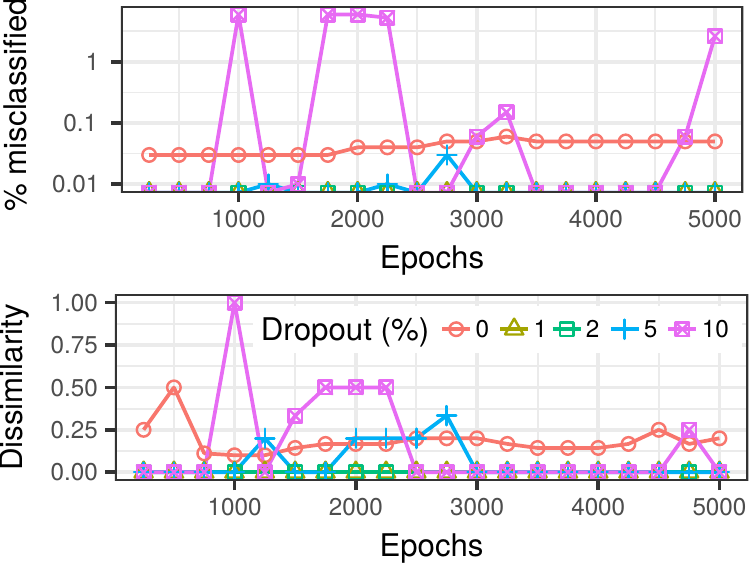}}
		\caption{Validation errors and dissimilarity measurement of 16-neuron GRU networks using dropouts. \label{fig:dropout}\vspace{-3mm}}
	\end{center}
\end{figure}

\textbf{Training set: }
We carried out a comparison between LSTM and GRU. The networks are trained with 400 and 1,000 examples and verified with 10,000 program paths. Misclassification percentages and dissimilarity ratios are shown in Figure~\ref{fig:lstmgru}.  
The lower values on the graphs are preferable. 
When dissimilarity ratio is zero, the framework is able to induce the expected rules.
Figure~\ref{fig:lstmgru}. With insufficient training data, the quality of knowledge extraction decreases noticeably.
The learning in both GRU and LSTM models improve as the training data set increased to 1,000.
The larger training set covers wider range of program examples, and thus, is better represent the target rules. 
Hence, more consistent extraction results are shown in the larger data set with a tradeoff of training time.

\textbf{LSTM versus GRU:} We have observed that GRU requires smaller training data than LSTM to successfully perform rule induction.
With 400 examples, GRU clearly out-performs LSTM as correct solutions cannot be extracted from LSTM models but can be extracted from GRU models with four neurons. 
With 1,000 examples, both LSTM and GRU can eventually learn the language at any size of hidden states with few exceptions in the early stage of training in LSTM.
GRU shows a better performance than LSTM as it has fewer parameter to be trained. 
Our results align with previous study~\citep{irie16} that GRU achieved slightly better results than LSTM with smaller networks.

\textbf{Regularizations: }
To reduce the overfitting, we tune our neural networks by adding a dropout layer to the input layer. 
We apply 0\%--10\% dropouts to GRU networks consisting 16 hidden states. The networks are trained with 400 examples. 
This model is overfit when no dropout is applied.
Regularization improves the consistency of rule extraction; 1\% and 2\% of dropout eliminate the overfitting problem as shown in Figure~\ref{fig:dropout}.
The learning is highly sensitive to dropout percentages as more noise is introduced.
The optimal dropout percentage is problem-oriented and may not hold true in different analysis.

\begin{figure}[tb]
	\begin{center}
		{\includegraphics{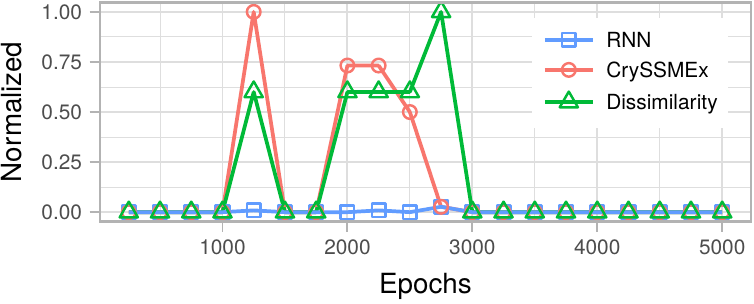}} \\
		\caption{Error propagations over the learning process in a 16-neuron GRU model using 5\% dropout .\label{fig:learningerrors}\vspace{-3mm}}
	\end{center}
\end{figure}

\textbf{Rule induction metrics: }
We evaluate learning accuracy of each pipeline stages using a test dataset. The output FSA is evaluated against the target FSA using similarity metric proposed in Section~\ref{sec:similarity}. 
An overlay of the similarity metric and machine learning accuracy are depicted in~\ref{fig:learningerrors}. 
Learning errors in RNN propagates to rule extractions. 
The trend of CrySSMEx's errors follows those of RNN's. Few exceptions, when the reasoning machine failed to extract RNN, generates steep spikes. 
The dissimilarity ratio reflects closely the errors in the rule extraction. 
We have observed that when an RNN model accurately predict the test set (i.e., $< 0.01\%$ inaccuracy), the rule extractor is able to extract the correct language.
The rule extractor is very fragile to error in the RNN.

\subsection{OpenJDK}
We further study the viability of our framework in a practical setup by studying caller-sensitive method (CSM) security vulnerabilities~\citep{cristina15} in OpenJDK~7 dataset.
A CSM is said to have a security bug if it is reachable from an untrusted source and  exists no permission checks before the CSM.
We construct ASM representation of OpenJDK's CFG using a datalog programming tool. Out-edges from CSM, permission checks and other basic blocks are labeled as \tts, \ttp~and \ttt~respectively.
CFG of OpenJDK contains several entry and exit points. Hence, we introduce an artificial start and end node that connected to the entry and exit respectively. The end node has an out-edge to the start node which allows the simulation of multiple OpenJDK library calls from untrusted sources. 
We apply the following assumptions: all valuables from untrusted sources are tainted, and the method is context insensitive. 
These may introduce false positive cases that may not reflect actual security bugs in OpenJDK. 
The ASM contains 1.5 million reachable  edges from untrusted sources. 
Trances of OpenJDK paths generate inputs for RNN.  
Each input is classified as secure or insecure according to the security specifications.
To avoid infinite loops ASM while tracing, we compute k-th shortest paths from the start node to CSM methods and then to the end node. 

\begin{figure}[tb]
	\captionsetup[subfloat]{labelformat=empty}
	\begin{center}
		\bgroup
		\setlength\tabcolsep{1pt}
		\renewcommand{\arraystretch}{0.9}
		\begin{tabular}{c| c c c c}
			\hline
			\diagbox{N}{Ep}	
			& 1000 & 2000 & 3000 & 5000 \\ \hline \hline
			
			4 &
			\subfloat[\scriptsize{$\Delta = 1.00$}]{\includegraphics[width = 0.45 in, height=0.7 in,clip]
				{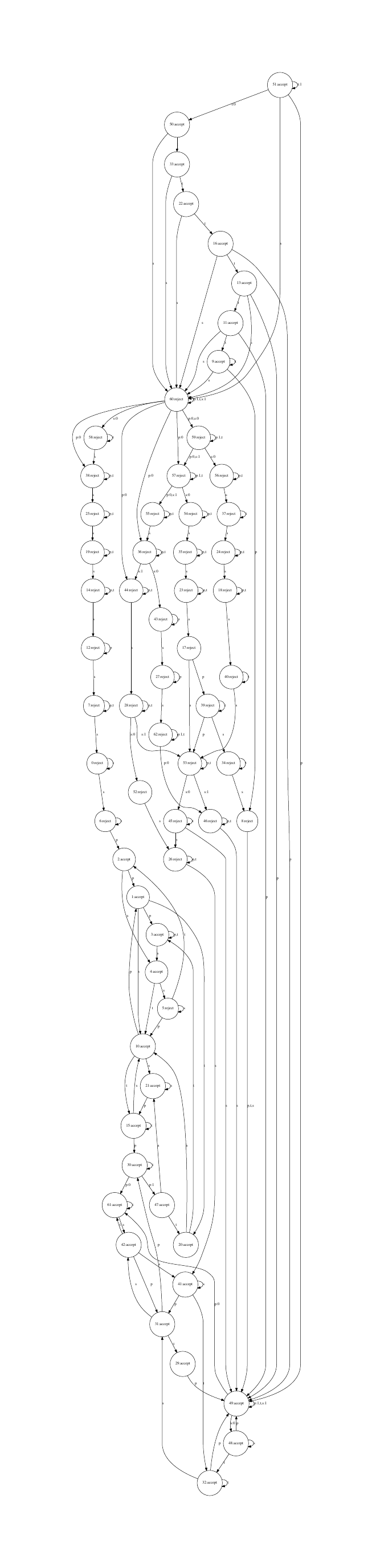}} &
			\subfloat[\scriptsize{$\Delta = 0.00$}]{\includegraphics[width = 0.7 in]
				{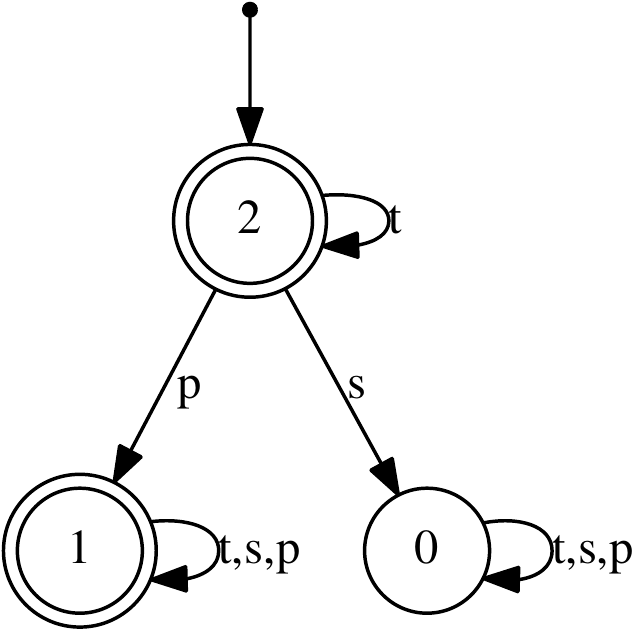}} &
			\subfloat[\scriptsize{$\Delta = 0.00$}]{\includegraphics[width = 0.7 in]
				{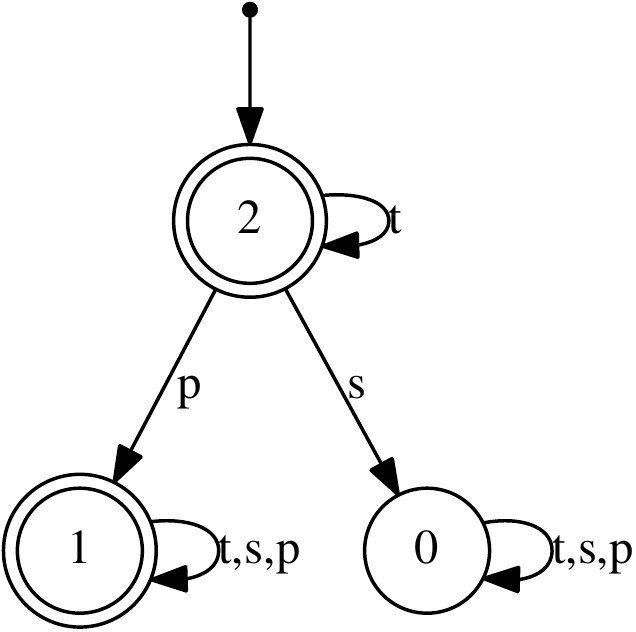}} &
			\subfloat[\scriptsize{$\Delta = 0.00$}]{\includegraphics[width = 0.7 in]
				{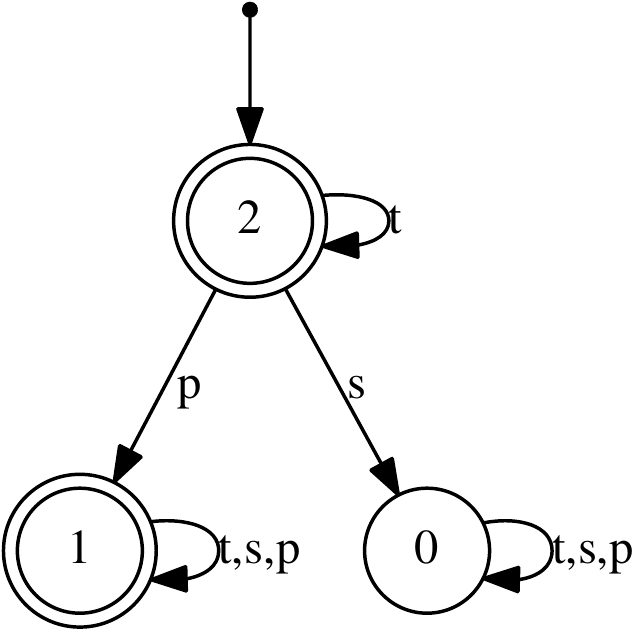}}\\ \hline
			
			8 &
			\subfloat[\scriptsize{$\Delta = 0.00$}]{\includegraphics[width = 0.7 in] 
				{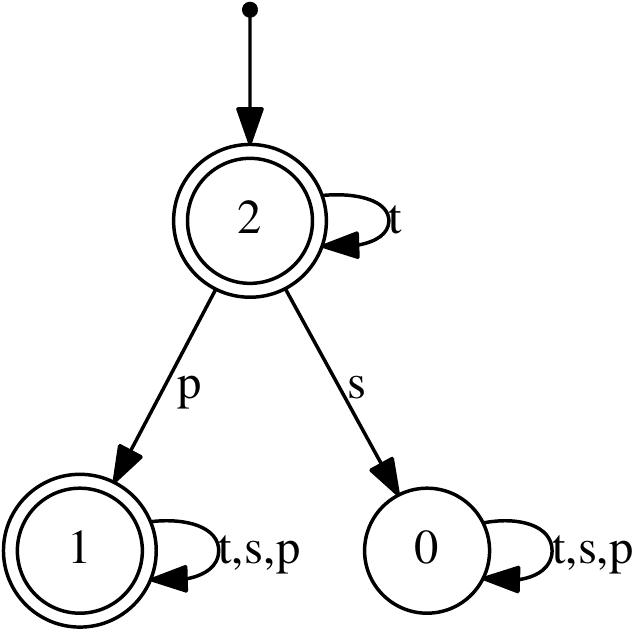}} &
			\subfloat[\scriptsize{$\Delta = 0.50$}]{\includegraphics[width = 0.5in, height=0.8 in]
				{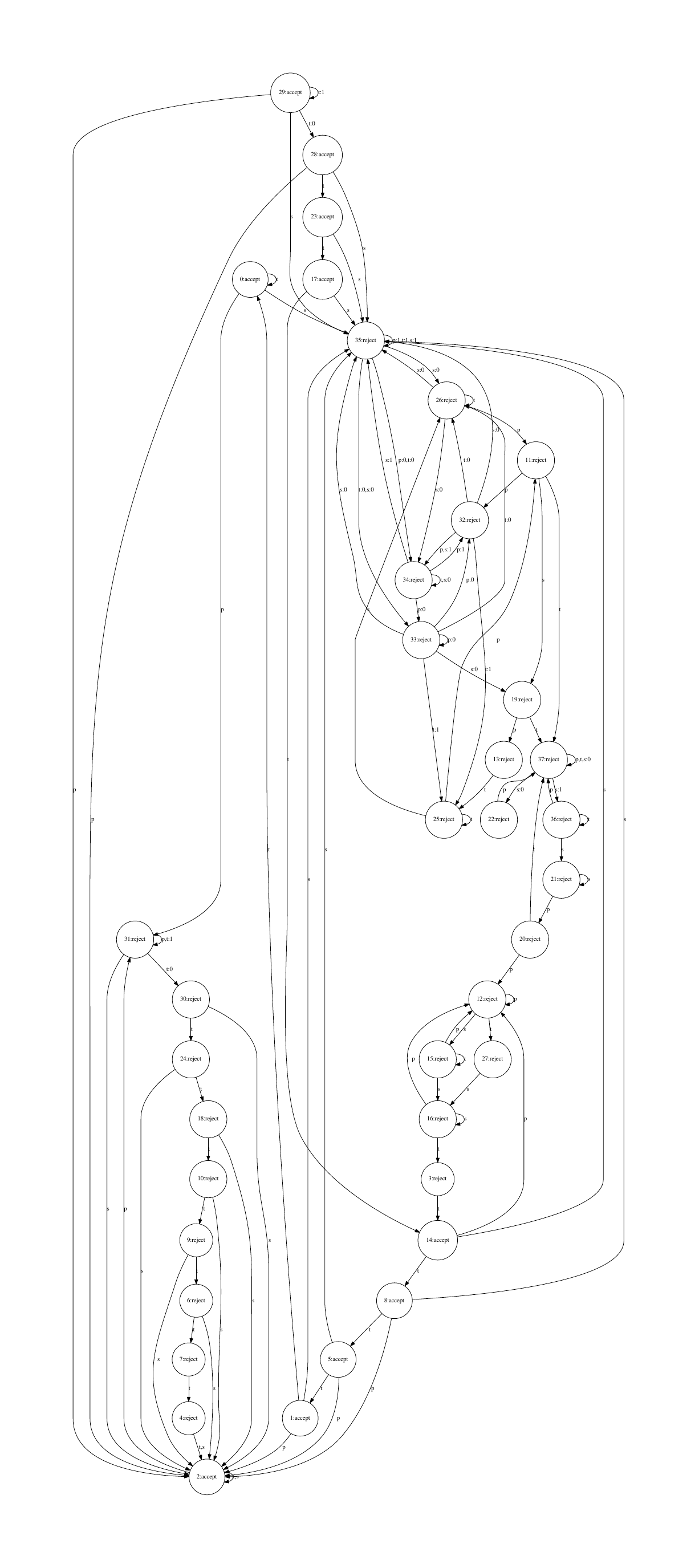}} &
			\subfloat[\scriptsize{$\Delta = 0.50$}]{\includegraphics[width = 0.45in, height=0.8 in] 
				{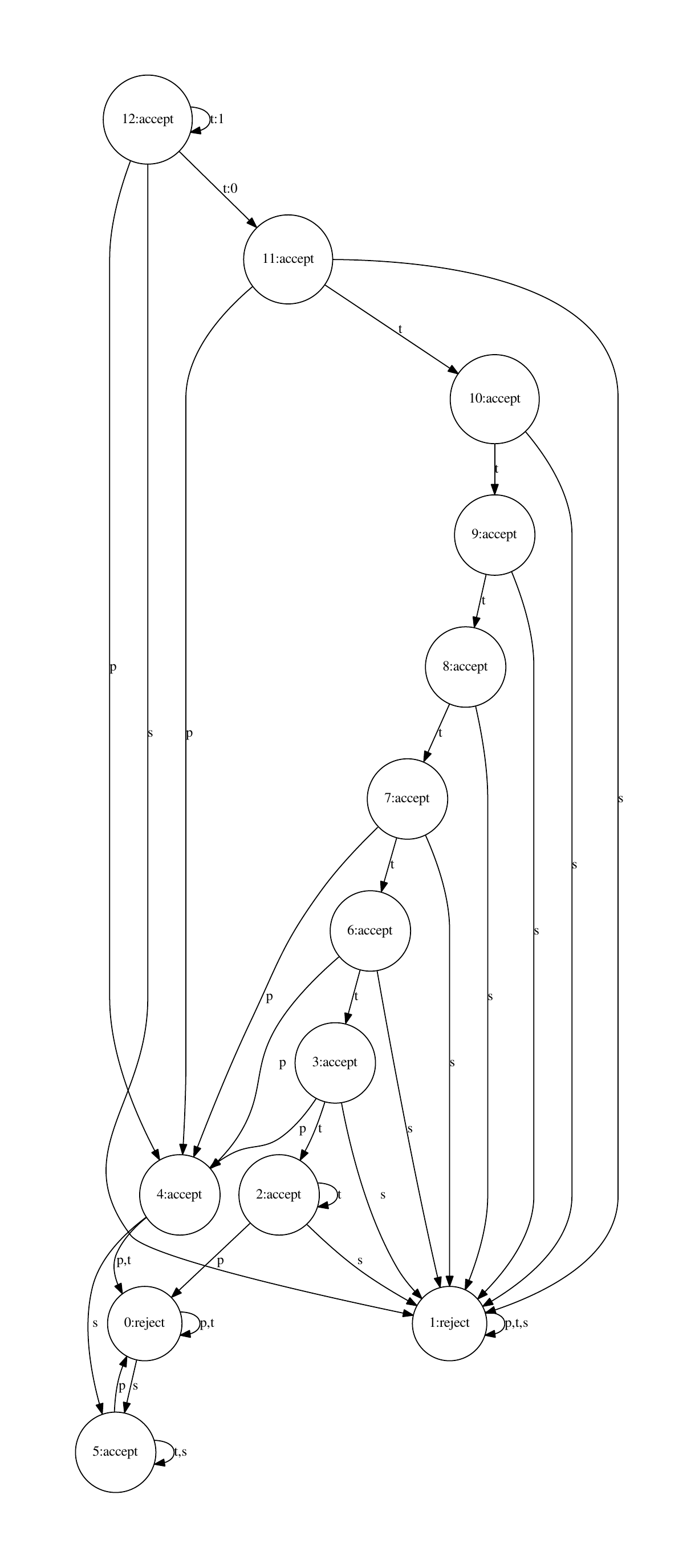}} &
			\subfloat[\scriptsize{$\Delta = 0.25$}]{\includegraphics[width = 0.45in, height=0.75 in]
				{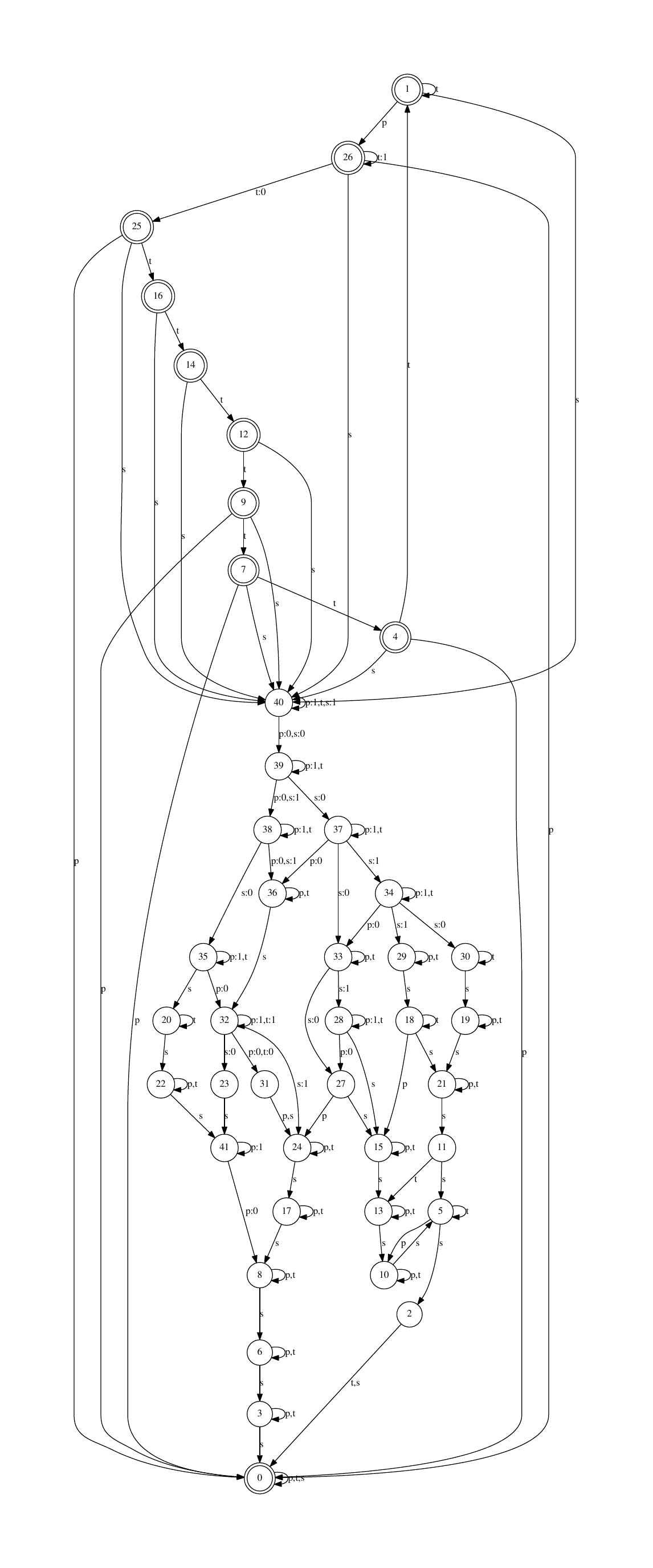}}\\  \hline
			
		\end{tabular}
		\caption{Rule extraction results from OpenJDK dataset\label{fig:learnedopenjdk}\vspace{-3mm}}
		\egroup
	\end{center}
\end{figure}

\begin{figure}[t]
	\begin{center}
		\begin{tikzpicture}[
		nnode/.style={draw, circle, fill=white, minimum size=1.2em}, 
		mat/.style={row sep = 1.2em,column sep=1em}, scale=0.9, every node/.style={scale=0.9}
		]
		
		\matrix[mat] (RAND){
			& \node [nnode,double, double distance=1pt, align=left] (a0) {0}; &\\
			\node [nnode,double, double distance=1pt, align=left] (a1) {1}; & & \node [nnode, align=left] (a2) {2}; \\
		};
		\node [above = 0.5 cm of a0] (initA){};
		\path[-latex]
		(initA) edge (a0)
		(a0) edge node [left,pos=0.4]{\ttp~$(0.404)$} (a1)
		(a0) edge node [right,pos=0.4]{\tts~$(0.268)$} (a2)
		(a0) edge[loop right] node {\ttt~$(0.328)$} (a0)
		(a1) edge[loop below] node {\ttt,\tts,\ttp~$(1.000)$} (a1)
		(a2) edge[loop below] node {\ttt,\tts,\ttp~$(1.000)$} (a2);

		\matrix[mat, right = 4em of RAND] (ACTL){
			& \node [nnode,double, double distance=1pt, align=left] (a0) {0}; &\\
			\node [nnode,double, double distance=1pt, align=left] (a1) {1}; & & \node [nnode, align=left] (a2) {2}; \\
		};
		\node [above = 0.5 cm of a0] (initA){};
		\path[-latex]
		(initA) edge (a0)
		(a0) edge node [left,pos=0.4]{\ttp~$(0.059)$} (a1)
		(a0) edge node [right,pos=0.4]{\tts~$(0.184)$} (a2)
		(a0) edge[loop right] node {\ttt~$(0.757)$} (a0)
		(a1) edge[loop below] node {\ttt,\tts,\ttp~$(1.000)$} (a1)
		(a2) edge[loop below] node {\ttt,\tts,\ttp~$(1.000)$} (a2);
		\node [below = 2em of RAND] {(a) Synthetic benchmark};
		\node [below = 2em of ACTL] {(b) OpenJDK};
		\end{tikzpicture}
		\caption{Markov chain\label{fig:markovchain}}
	\end{center}
\end{figure}
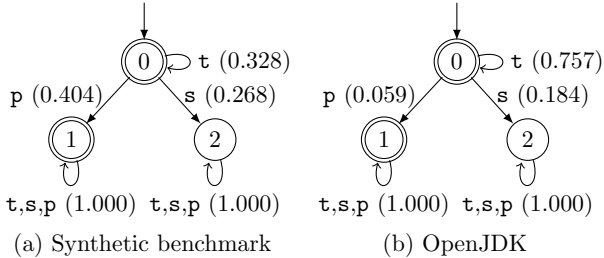

\begin{figure}[tb]
	\begin{center}
		{\includegraphics{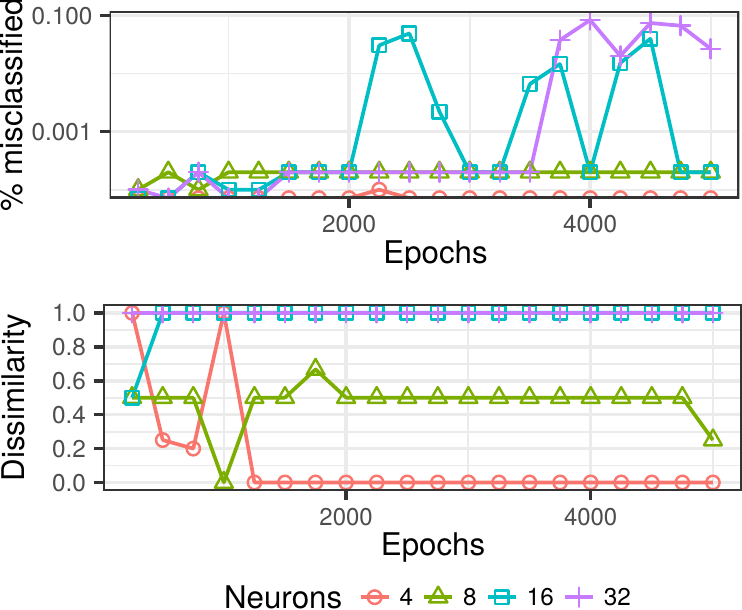}} \\
		\caption{Error in classification and dissimilarity metric of OpenJDK.\label{fig:openjdk_class}\vspace{-3mm}}
	\end{center}
\end{figure}

Figure~\ref{fig:markovchain} shows that program paths in OpenJDK heavily bias with a low probability of having permission checks, \ttp,  whereas the synthetic benchmark is nearly uniform distributed. 
The low probability limits learning ability of RNN. The imbalance distribution can be mitigated by improve a loss function or improve positive to negative sample ratios. The former option reduce the generality of the framework as the updated loss function becomes application specific. The latter option is feasible as the ASM of OpenJDK is known. We restrict 30\% of training data are positive samples. With improved distribution, symbolic rule induction is feasible.

Similar to the synthetic experiment setup, the GRU model is trained with 1,000 training data and 1\% dropout. Figure~\ref{fig:learnedopenjdk} shows the induced FSA of RNN using four and eight neurons. We have conducted the experiments for 16 and 32 neurons, but no FSA could be extracted within the time limit.
Figure~\ref{fig:openjdk_class} depicts learning accuracy in RNN and similarity of the target FSA.  
Comparing to the synthetic experiment (see Figure~\ref{fig:lstmgru} with 1,000 training data), it is more difficult, but possible, to induce the correct symbolic rules. 
The lower degree of learning is the result of certain behavior in actual programs. For instead, there exists $\frac{1}{3}$ chance of \ttp~followed by \tts~ in the synthetic benchmark but does not hold true in OpenJDK. The patterns are encoded in RNN and later results in erroneous rules.
Moreover, we have observed that the correct symbolic rules can only be learned from RNN models with no misclassifications

In short, learning symbolic rules from real-world dataset is feasible with sufficient training data and fair distribution of example. The existing rule extraction tool is sensitive to error in RNN models. Hence, high accuracy in RNN training is required for a successful induction.

\section{Related work}
\label{sec:relwork}

Grammatical inferences have been widely studied since the 1970s. 
A thorough survey on grammatical inferences was conducted by ~\citet{higuera10}. 
A \textit{active learning} solution to regular language inference has been  introduced by Angluin~\citep{angluin87}. 
The algorithm constructs a regular language by using membership queries asking an oracle, who answers if a string belong to the target language, and supply  counterexamples, otherwise. 
PAC learning model~\citep{valiant84} approximates a DFA with high probability from random examples with an assumption on an example distribution. Learnability of PAC under any distributions is considered to be too restrictive. 
\citet{muggleton14} proposes using abduction in inductive logic programming for grammar inferences. 

The early noticeable works in this area~\citep{giles92,cleeremans89,watrous91} discover FSA from a simple recurrent network with an assumption that after sufficient training, the network's state space clearly diverges and clusters of the state vectors correspond to a state of an FSA. 
\citet{kolen94} gives a thorough examination to the past techniques and address two critical problems. First, sensitivity to the initial conditions which may give fault illusion of the outcome.
Second, trivial changes in observation strategies can cause one to induce behavior. An RNN state control behavior of current and future events. Thus, a small distance in the state space may give larger impact to the network functionality than a large distance.

Jacobsson~\citeyearpar{jacobsson06} has shown that \cryssmex{} can extract FSA from a context free language, $a^{n}b^{n}$. Also Frank and Jacobsson~\citeyearpar{frank10} applied the rule extractions on echo state networks that learn word prediction.
With renewal interests of deep-learning in the resent decade, RNN models, including LSTM and GRU have been greatly refined. Rule extraction algorithm has yet been evaluated on recent RNN models.

Mou et al.~\citeyearpar{mou16} introduces a tree-based convolutional neural network (CNN) for analysis of abstract syntax tree. Each tree node is encoded in a vector with a numeric representation. By learning ASTs of programs from open judge system, the tree-based deep-learning can classify programs by features.
Montavon et al.~\citep{montavon17} proposed deep Taylor decomposition that explains contributions of input data to the classification decision. The algorithm traverses a trained network backward from the output layer to reveal how decisions were made. 
\citet{zaheer2016} develop a static analyzer induction framework for a toy language. The framework learns program properties through LSTM and extracts rules using differentiable set.  \citet{bielik2017} take a different approach. They induce static analyzers using synthesis algorithm then generalize learn analyze by querying counterexample from an oracle.

Recently, there has been interested in learning program analysis using probabilistic models.
Raychev et al.~\citeyearpar{raychev15} propose a new framework that uses conditional random fields to predict program facts. The framework is trained by \textit{Big Code},  programs in large code repositories, such as GitHub. This prediction engine is used in JSNice to predict variable names and types in context of JavaScript.
Long et al.~\citeyearpar{long16} a patch generation system using probabilistic model, which learns the correct code from human-written patches to fix new program bugs.

\section{Conclusion}
\label{sec:conclusion}
\vspace{-2mm}
While neural networks and deep learning provide high predictive accuracy,
the induced models are not ``explainable''. Our machine learning pipeline 
extracts a reasoning machine from an RNN to make models explainable.
It includes (i)~modeling program behavior from examples and (ii)~extacting symbolic rules, in form of a regular language, from the RNNs using a probabilistic finite state machine. 
The output program analyzers analyze potential errors in other programs. We present a sampling-based similarity measurement between two regular languages to evaluate the robustness of the learning pipeline. Our case study shows a successful induction of the security analzyer using LSTM and GRU models when sufficient training data and fair distribution of program paths are provided. However, we have observed that the rule extractions from RNNs is very sensitive to model accuracy. Improvements in rule extraction is needed to induce more complex analysis.

\bibliographystyle{abbrvnat}
\bibliography{ml4spa}
\end{document}